# High School Class for Gifted Pupils in Physics and Sciences and Pupils' Skills Measured by Standard and Pisa Test

**G. S. Djordjevic**[a] and **D. Pavlovic-Babic**[b]

[a]*Department of Physics, University of Nis, P.O.Box 224, Nis, Serbia*

[b]*Institute for Psychology, Faculty of Philosophy, Cika Ljubina 18-20, Belgrade, Serbia*

**Abstract.** The ``High school class for students with special abilities in physics`` was founded in Nis, Serbia (www.pmf.ni.ac.yu/f_odeljenje) in 2003. The basic aim of this project has been introducing a broadened curriculum of physics, mathematics, computer science, as well as chemistry and biology. Now, six years after establishing of this specialized class, and 3 years after the previous report, we present analyses of the pupils` skills in solving rather problem oriented test, as PISA test, and compare their results with the results of pupils who study under standard curricula. More precisely results are compared to the progress results of the pupils in a standard Grammar School and the corresponding classes of the Mathematical Gymnasiums in Nis. Analysis of achievement data should clarify what are benefits of introducing in school system track for gifted students. Additionally, item analysis helps in understanding and improvement of learning strategies' efficacy. We make some conclusions and remarks that may be useful for the future work that aims to increase pupils` intrinsic and instrumental motivation for physics and sciences, as well as to increase the efficacy of teaching physics and science.

## INTRODUCTION

In the countries of former Yugoslavia, classes for students with special abilities have a long and successful tradition. Despite some improvement that has been made through last few years, today three fundamental problems still characterize the teaching at the schools in Serbia: obsolete equipment, obsolete education concepts and insufficient motivation of teachers (small payroll). One of the consequences of this situation is a very small number of students of the natural sciences and engineering sciences at the university. Moreover skills of pupils in using methods and tools developed in physics and other sciences seem to decrease at the same time with a new revolution in science and technology in developed countries.

The main goals of the project "GRAMMAR SCHOOL CLASS FOR STUDENTS WITH SPECIAL ABILITIES IN PHYSICS" are to offer a high-quality education, to give gifted pupils a perspective for continuing with high-quality education and to convey initiative and enthusiasm [1,2]. These goals are to be reached by the following measures: (i) focus on the natural science, on physics in particular, (ii) provision of basic laboratory equipment and PCs (virtual experiments and Internet access), (iii) close collaboration with the University (Host teaching by docents, assistants and project guests, mentors from the university), (iv) close collaboration with similar projects in EU and Eastern Europe, (v) more intensive foreign languages teaching (especially English), for details see [2,3].

The authors of the curricula and project have faced a lot of problems in implementation of the project in its basic form during last three years. However, one of the most important aims has been permanent evaluation of the pupils in the ``new class`` and comparison of their results with pupils educated in the standard and ``mathematical`` classes in Serbia. In this paper we present some results in a brief form.

# EVALUATION IN PHYSICS (2003-2006)

Let us remind, briefly, on the main results in the previous phase of the continuous evaluation of the ``Special Class``, i.e. Cirrciula.

Results of two tests in physics made in October 2005 and May 2006 are given in the Table 1. There have been five groups of pupils (1. Special class for "physicists" in Nis (9 pupils), 2. Special class for "mathematicians" in Nis (7), 3. Standard grammar class in Nis (20), 4. Special class for "mathematicians" in Belgrade (17) and 5. Special class for "mathematicians" in Novi Sad (7). All pupils worked out the same test with 20 questions (in total 100 points) and 2 problems (in total 50 points). At this stage we measured abilities of pupils only in physics and mainly in the first class. The differences in syllabus in physics are so big in the second and third year that comparison of results is sensible just after the end of the grammar school, i.e. after $4^{th}$ year.

In the table bellow we present results of the third generation - pupils born in 1990. The other results will be presented elsewhere. In the first column we denote the corresponding class and the numbers of pupils that took part in both testing. In the following columns one can see their records in test the questions and solving problems in percents.

**TABLE 1.** Results of the third generation

| | Questions 1 (%) | Problems 1 (%) | Total 1 (%) | Questions 2 (%) | Problems 2 (%) | Total 2 (%) |
|---|---|---|---|---|---|---|
| "Physicists"-Nis (9) | 52,67 | 22,4 | 42,58 | 71,11 | 27,78 | 56,67 |
| „Mathematicians"-Nis (7) | 35,00 | 0,00 | 23,33 | 64,86 | 4,29 | 44,67 |
| Standard class-Nis (20) | 42,60 | 0,00 | 28,40 | 61,40 | 0,00 | 40,93 |
| "Mathematicians"-BG (17) | 67,06 | 0,71 | 44,94 | 68,29 | 29,18 | 55,25 |
| "Mathematicians"-NS (7) | 69,14 | 8,29 | 48,86 | 81,86 | 31,43 | 65,05 |

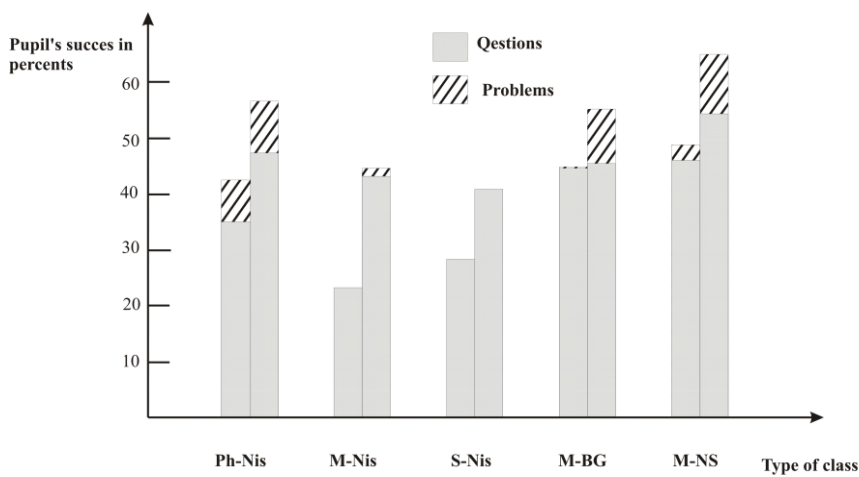

**FIGURE 1.** Pupil's success in solving test questions and problems. Ph-Nis, M-Nis, S-Nis, M-BG, M-NS.

# EVALUATION IN PHYSICS (2006-2008)

There were several tests in physics during that period (2006-2008). It should be noted that every of the first 4 generations of ``physicists`` has been followed and evaluated, as well as the corresponding group of ``mathematicians`` and at least one ``control`` group of pupils in a Grammar school - ``standard Curricula``, mostly in Nis. Presentation of the contemporary results and analysis would need much more ``room`` than we have at our disposal in this paper. However, we will present results concerning the first generation of ``physicists`` and groups of pupils we evaluated permanently through their 4 years education it the Grammar schools.

Let us consider results of two independent tests made by different groups. An independent group of evaluators nominated by the Serbian Ministry of Education both made in spring 2007. The first test was prepared by the ``external`` team of the evaluators. There were 4 groups of questions and problems in physics, with increasing complexity, from 1 to 4. The results are presented in the Table 2. All pupils were in the 4$^{th}$ grade and they took part in the first testing, as well as in all other evaluations started in September 2003.

| TABLE 2. Results of the first generation – 4$^{th}$ grade | | | | | |
|---|---|---|---|---|---|
| **Level of complexity** | 1 (%) | 2 (%) | 3 (%) | 4 (%) | Total (%) |
| "Physicists"-Nis (9) | 93,3 | 95,6 | 69,4 | 69,66 | 85,68 |
| „Mathematicians"-Nis (8) | 94,6 | 89,7 | 71,1 | 37,5 | 74,26 |
| Standard class-Nis (25) | 82,4 | 23,3 | 4,7 | 0,0 | 34,47 |

The second test was prepared by the permanent group of evaluators form Nis. This test was similar to the previous ones, slightly improved and also with classified problems in 5 categories. The results were summarized in Table 3.

| TABLE 3. Results of the first generation – 4$^{th}$ grade | | | | | | |
|---|---|---|---|---|---|---|
| **Level of complexity** | 1 (%) | 2 (%) | 3 (%) | 4 (%) | 5 (%) | Total (%) |
| "Physicists"-Nis (7) | 100,0 | 87,2 | 58,33 | 100,00 | 71,43 | 37,93 |
| „Mathematicians"-Nis (13) | 84,62 | 47,92 | 32,05 | 43,59 | 25,64 | 19,81 |
| Standard class-Nis (11) | 72,73 | 46,78 | 20,45 | 3,03 | 0 | 14,32 |

Numbers in the small brackets denotes numbers of the pupils form the corresponding group who took part in the particular evaluation. Despite a relatively small number of pupils and ``fluctuation`` of some pupils, when absence of a few best or less good form the group can change the total score. It is obviously that ``physicists`` after 4 years spend in the special class with the new curricula have excellent results, and that they abilities and skills in physics are significantly better than ``mathematicians`` who were selected at the beginning of his high school study (``physicists`` in this generation were not because they could enroll on the basis of their wishes). Let us remind on the results of the same classes at the very first testing, without further comments.

| TABLE 4. Results of the first generation – 1$^{st}$ grade | | | |
|---|---|---|---|
| **Level of complexity** | 1 (%) | 2 (%) | Total (%) |
| "Physicists"-Nis (17) | 28,6, | 4,47 | 21,69 |
| „Mathematicians"-Nis (16) | 34,69 | 7,06 | 27,83 |
| Standard class-Nis (27) | 32,01 | 0,00 | 21,34 |

## PISA TEST

Trying to collect more valuable and comparative data about achievement of pupils and achievement correlates, we included OECD/PISA (Programme for International Student Assessment) science competencies tests in the evaluation schema. The main intent was to establish a model of monitoring and external evaluation of achievement focused on functional knowledge, which enable us  a) to compare the quality of knowledge in domain of physics and, even more, of science, to internationally recognized criteria (key competencies); b) to document  the well-being of schooling in special teaching program, e.g. comparing their achievement with classic course, and c) to understand in what extent higher order thinking is supported by different types of schooling.

PISA 2006 defines scientific literacy in terms of an individual's scientific knowledge to identify questions, to acquire new knowledge, to explain scientific phenomena, and to draw evidence-based conclusions about science-related issues. In addition, pupil shows the understanding of the characteristic features of science as a form of human knowledge and enquiring. For detailed description see [4]. In other words, PISA is not limited to measure pupil's acquisition of curricula or specific science content, but his/her capacity to identify and understand science related issues as well as the capacity to interpret and apply science evidence in order to solve problems and make decisions in real-life situations [5].

## Sample

All pupils in special classes for physics in the 4[th] and the 1[st] grade (11, 10 pupils, respectively) were included in testing. In addition, in order to compare data, special math class grade 4[th], Grammar school the 4[th] and the 1[st] grade, were included as well (14, 11, 32 pupils respectively). Also, we use PISA 2006 national database to compare achievement data.

We are aware that all conclusions we drawn, based on comparison between this and national sample, are uncertain due to small size of the selected sample, as well as to the age difference (PISA pupils are mostly 1[st] grade pupils).

Testing took part at the end of the school year (May 2007), i.e. at the beginning and at the end of the secondary schooling. Every student was given a 2-hours paper-pencil science tests.

## Results

### Science Competencies: Average Achievement

The science scale was constructed to have a mean score among OECD participating countries of 500 points, and standard deviation of 100 points, which means that it is expected to have about two thirds of pupils scoring between 400 and 600 points. Table 4 shows the average achievement of OECD countries pupils and Serbian pupils in PISA 2006.

**TABLE 4.** Average performance – PISA 2006.

| Referent sample | Mean |
|---|---|
| OECD countries | 500 |
| Serbia, the whole sample | 436 |
| Serbia, Grammar school pupils | 501 |

In short, results show us that, as a whole, Serbian education system performs under the expected average. Knowing that one year of schooling adds about 40 points on this scale, we can say that Serbian education has lost (or has spent on nothing) one and a half school year out of 9 years (which is the duration of formal schooling of PISA pupils in the moment of testing). Grammar school pupils are at the level of their OECD counterparts, but this is the whole OECD sample, which includes all groups of pupils, while in Serbia, in Grammar schools we have pupils positively selected by their school competencies.

When looking at Figure 2. we can notice that "physicists" grade 4[th] is slightly above this average achievement but here we have 3 school years older pupils. On the other hand, the achievement of two remaining special classes is far above the average (more then two standard deviation). Those results are respectful even knowing that we are talking about two relatively small groups of pupils. Additional analysis helped us to understand better the nature of these outstanding results.

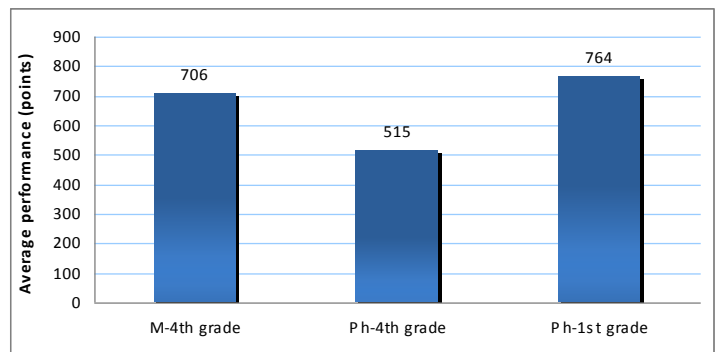

**FIGURE 2.** The average performance of gifted pupils on science scale.

*Science Competencies: Levels of Achievement*

Pupil performances on science are grouped into six proficiency levels. This grouping was taken on the basis of substantive considerations relating to the nature of the underlying competencies. Each pupil is posed in the highest level in which she or he can answer the majority of tasks. So, levels 1 and 2 show what is the critical baseline for science competencies. On the top of the scale (proficiency levels 5 and 6) we have pupils enabled to solve complex problems, related to scientific knowledge and understanding scientific data, which require several interrelated steps, as well as the use of critical thinking and abstract reasoning [6]. Following graphs show some comparative data about the distribution of pupils in these two top levels of performance.

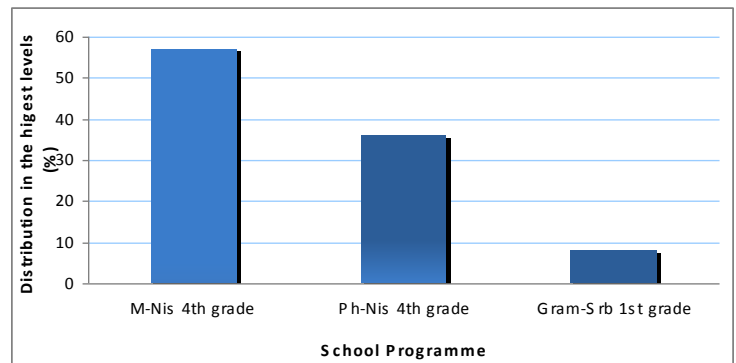

**FIGURE 3.** The percentage of pupils in the highest levels of performance (4th grade – gifted pupils and PISA Serbia Grammar school sample)

Figure 3. shows us that high performance is very rare among 1st grade pupils of Grammar schools in Serbia, less than 10% of them can successfully solve the most complex PISA tasks. At the same time, more than half of gifted mathematicians are able to perform on these levels. But, the most interesting data is referring to the distribution of performance in the group of "physicists". While this group of pupils performs, in average, at the same level as the Grammar school students, here we can see higher concentration of top performances in this group than in Grammar school group (almost three times more). Still, we should keep in mind the fact that we speak about pupils of different ages and, consequently, of different length of schooling, as well as the very small groups of pupils in special classes. So, let us see what the result of comparison between the 1st grade gifted pupils and their counterparts in Grammar schools is (Figure 4.)

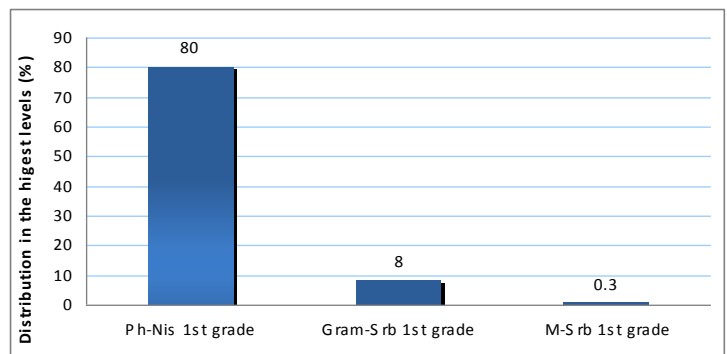

**FIGURE 4.** The percentage of pupils in the highest levels of performance (the 1st grade – gifted pupils, PISA Serbia Grammar school and PISA Serbia whole sample)

As we can see, almost all pupils in this class have the highest ranking scores. Talking in language of competencies, these pupils demonstrate advanced scientific thinking and reasoning, and they are able to solve complex and unfamiliar scientific or technological situations.

We can draw some policy-oriented implications on these results. First of all, it is obvious that groups of gifted pupils, especially those in 1$^{st}$ grade, deserve and/or need very carefully designed curricula which will engage their already built competencies, as well as enable them to develop their capacities even more. Also, it is very important for them to have well-trained teachers sensitive enough to follow their educational needs. Then, we can say that demonstrated thinking skills have transfer effects on other (related) domains, so we can expect them to be high achievers in other subjects, as well.

## CONCLUSION

Let us note that this program of evaluation of the curricula and its implementation, even though it has been focused just on physics at this stage is quite nontrivial. There are 5 groups of pupils in 3 different cities and 4 schools. About 10 people are included in this evaluation practically on a complete voluntary basis. Finally syllabus on physics is similar just in the first year and after that rather different until the end of the grammar school when they have the same "corn" in physics.

Let us denote that pupils from the standard class are not able to solve problems (their records in solving problems tend to zero in all three generations). The "physicists" show slightly better improvement in physics, and continually good records in solving problems. Let us note that the second test in Novi Sad was done just by seven best pupils chosen by their school. It is worth to note that the new class and program "for physicists" has attracted better pupils and that number of pupils is increasing 7, 11 and 15 in 2$^{nd}$, 3$^{rd}$ and 4$^{th}$ (the newest) generation. It can be explained by attractive curricula, a lot of guest lecturers, additional laboratory work, excursions, some support in books and awards.

## ACKNOWLEDGMENTS

This work is partially supported by UNESCO-ROSTE grant No. 8759145, and UNESCO-BRESCE grants No. 8758346 and 8759228, as a part of the Southeast European Network in Mathematical and Theoretical Physics (SEENET-MTP), www.seenet-mtp.info . We would like to thank all the colleagues from Nis, Belgrade and Novi Sad who take part in this six years long period of permanent evaluation. We are exceptionally thankful to Lj. Nesic, T. Misic and Lj. Kostic-Stajkovic, for preparation of many tests and their evaluation. It is our pleasure to thank D. Dimitrijevic and J. Stankovic for their continual support in preparing test materials and their realization.